\documentclass[12pt]{article}
\usepackage{amsmath, amssymb, graphicx, authblk}
\usepackage[a4paper, margin=1in]{geometry}
\title{Is {\it E. coli} good at chemotaxis?}
\author{Robert G. Endres}
\affil{Physics of Life Network of Excellence \& Department of Life Sciences, Imperial College London, UK}
\date{}

\begin{document}

\maketitle

\noindent {\bf Disclaimer:} This is an early pre-acceptance version of a News \& Views article later accepted by Nature Physics \cite{Endres2026NatPhys}. The published version, with a 6-month embargo, differs in title, wording, and layout. Specifically, the historical reference to Ancient Greece and the art work were removed to align with a more contemporary publishing style. Copyright © Springer Nature.\\ \ \\


\noindent{\bf Bacteria seem masters of chemotaxis, yet new work suggests otherwise. Henry Mattingly and colleagues argue cells squander most sensory information, making chemotaxis surprisingly far less efficient than Berg and Purcell’s physical limits allow.}\\

\noindent In Greek mythology, Sisyphus is condemned to push a boulder uphill for eternity, only to see it roll back down. Bacteria face a similar fate: every glimpse of a chemical gradient is blurred by noise and diffusion (see Fig. 1 for an illustration). Yet, unlike Sisyphus, they make progress — transforming randomness into directed motion. In new work, Henry Mattingly and colleagues integrated an information-theoretic framework with experimentally determined parameters to test how close bacteria come to the physical limits of sensing \cite{mattingly2024natphys_submitted}. They reported that cells fall well short, losing most information during signal processing and converting only a fraction into chemotactic drift.\\

\begin{figure}[t]
\centering
\includegraphics[scale=1.1]{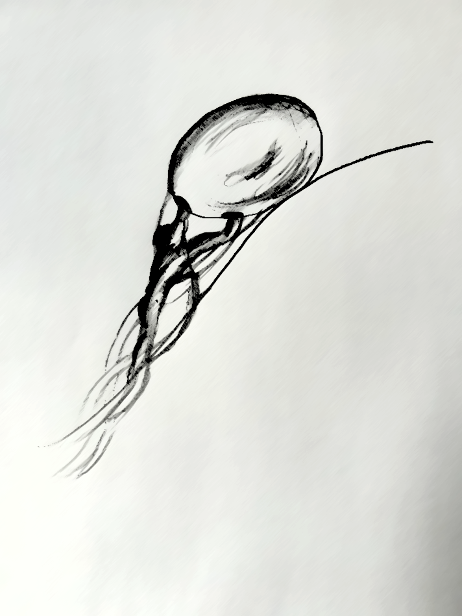}
\caption{{\bf Sisyphus at the microscale.} Like the mythical king, a bacterium struggles uphill—pushing against molecular noise and random reorientations. Unlike Sisyphus, however, it manages to steadily inch its way up the gradient. Hand drawing by Robert Endres.}
\label{fig:fig1}
\end{figure}


\noindent In a seminal 1977 paper, Howard Berg and Edward Purcell asked how accurately such a small device could measure its chemical environment \cite{berg1977biophysj}. They treated the cell as a sphere, able to count diffusing molecules and estimate concentration (assuming cells also know their volumes and basic arithmetic). They showed that the uncertainty of concentration estimates decreases with faster diffusion, larger cell size, and longer time for averaging. Their analysis established a physical limit for sensory accuracy, often referred to as the Berg–Purcell limit \cite{AquinoWingreenEndres2016ThinkAgain}.\\


\noindent Bacteria use a navigation strategy based on "biased random walk": alternating straight swims ("runs") with random reorientations ("tumbles"). Because bacteria are too small to sense reliable concentration differences across their length before rotational diffusion scrambles orientation, they rely on memory to compare concentrations along a run. If conditions improve, runs are extended; if not, cells tumble sooner. Berg and Purcell showed that such temporal comparisons can succeed, provided the gradient signal is stronger than the uncertainty of measurement.\\

\noindent Nearly fifty years later, the problem has been revisited with powerful new tools \cite{mattingly2024natphys_submitted,mattingly2021natphys}: microscopy that reveals signaling inside single cells, information theory that quantifies information flow and loss, and simulations that explore behavior in noisy environments. These advances allow researchers to ask a sharper question: do bacteria actually operate at the physical sensing limits set out by Berg and Purcell, or does information loss in their biochemical pathways prevent them from reaching this bound? The surprising answer is that performance is far below the limit \cite{mattingly2024natphys_submitted}.\\

\noindent Henry Mattingly and colleagues present an information-theoretic analysis of {\it E. coli} chemotaxis, asking how efficiently cells convert a tiny, time-varying cue in log-concentration (the relative gradient they encounter with time when swimming) into an internal response that can bias motion \cite{mattingly2024natphys_submitted}. They define two dimensionless signal-to-noise measures: one for ligand-arrival statistics at the cell surface ($\gamma_r$) and one for the downstream kinase activity ($\gamma_a$). From these, they compute corresponding steady-state information rates from signal to arrivals and from signal to activity, and use their ratio, $\eta=\gamma_a/\gamma_r$, as a transmission efficiency. Strikingly, they report efficiencies on the order of about 1\%, implying that most of the physically available information is lost before reaching the decision layer.\\

\noindent Why so low? The authors argue that substantial information is dissipated along the pathway—through finite bandwidth, adaptation, and internal biochemical noise—so cells may not operate at the physical limit even if the environmental signal itself is measurable. To ground parameters, they combine single-cell FRET measurements (to obtain the kinase response kernel and internal noise level) with known swimming statistics (speed, baseline tumble rate, persistence, rotational diffusion) to set the motor-behavior model. They also note phenotypic variability: some cells transmit more than others, yet even the best performers remain well below the physical bound in their framework.\\

\noindent Finally, they run controlled simulations that couple the measured pathway to a run-and-tumble kinematic model, and compare to an idealized “arrival-limited” cell that could exploit all arrival information. In those head-to-head tests, the ideal cell achieves a few-fold larger drift—underscoring their central claim: bacterial chemotaxis appears information-limited, not set by the classic sensing bound \cite{mattingly2024natphys_submitted,mattingly2021natphys}.\\

\noindent Cells may lose most of the information available at their surface, but they may not need more — at least that is what earlier work has suggested. Berg and Purcell noted that chemosensing is inherently robust: even 1\% receptor coverage is sufficient to approach the physical limit within a modest factor \cite{berg1977biophysj}, while Celani and Vergassola later argued that cells adopt a generalist’s maximin strategy that secures the highest minimum uptake of chemoattractants for any concentration profile \cite{Celani2010PNAS}. More closely related to the information-deterioration problem raised by Mattingly {\it et al.}, Brumley and colleagues fitted a drift model to experimental data and found only a few percent variability across wide-ranging noise levels \cite{Brumley2019PNAS}. Together, these results suggest that bacterial chemotaxis functions effectively even when information flow is inefficient.\\

\noindent Others have argued, however, that information optimisation may not be the whole story. Noise-induced fat tails in run-time distributions can produce Lévy-like walks, which favour long-term exploration and dispersal in patchy environments rather than precise short-term chemotaxis \cite{TuGrinstein2005PRL,Matthaus2011PLOS}. Biochemically detailed simulations will be essential to test this robustness further \cite{Micali2017BJ}, since the work of Mattingly {\it et al.} relies primarily on information-theoretic formulations rather than fully independent mechanistic simulations.\\

\noindent The study is compelling though, pushing the limits of what information theory and experiments can reveal together. Yet much remains to be understood about the diversity of chemotaxis strategies, even in {\it E. coli}, one of the best-characterised organisms in biology. Future directions include linking temporal and spatial sensing as cell size and speed vary across species \cite{WanJekely2021PRSB}, and testing robustness of chemotaxis in realistic ecological settings. With papers like this one, the field is finally asking the right quantitative questions. And remember: unlike Sisyphus, bacteria succeed.\\

\bibliographystyle{unsrt}

\bibliography{references}

\end{document}